\documentclass[12pt,preprint]{elsarticle}
\journal{Physics Letters B}

\usepackage[T1]{fontenc} 
\usepackage{framed}

\usepackage{url}
\usepackage{axodraw}

\usepackage{latexsym}
\usepackage{amssymb}
\usepackage{amsfonts}
\pdfoutput=1

\usepackage{amsmath}
\usepackage[caption=false]{subfig}

\usepackage{breqn}

\newcommand{\ep}{\epsilon}

\newcommand{\dr}{{\rm d}}

\newcommand{\nn}{\nonumber}

\allowdisplaybreaks[4]
 
\begin{document}
\begin{frontmatter}

\title{Simplifying the large mass expansion}
 
\author[SINP]{V.A.~Smirnov}
\ead{smirnov@theory.sinp.msu.ru}
 
\address[SINP]{Skobeltsyn Institute of Nuclear Physics of Moscow State University,\\  119992 Moscow, Russia}
 
\begin{abstract}
It is shown how the well-known large mass expansion can be simplified to obtain more terms of the expansion
in an analytic form. Expanding two-loop four-point Feynman integrals which contribute to the process $H \to ggg$ is used as an example.
\end{abstract}
\end{frontmatter}
\newpage
%

\section{Introduction}
 
The large mass expansion is known for more than forty years. It was successfully applied in numerous calculations.
The large mass limit, as well as the off-shell large momentum limit, are examples of
limits typical of Euclidean type which are characterized by considering some external momenta as large or small in the
Euclidean sense. Formally, an external momentum $q_i$ is characterized as small if it is scaled as $q_i\to \rho q_i$ 
with $\rho\to0$, and all the other external momenta are called large. So, in the large mass limit, all the external momenta are small in this sense and some of the masses are large.
The behaviour of a given Feynman integral in the large mass limit can be described 
\cite{Chetyrkin:1988zz,Gorishnii:1989dd,Smirnov:1990rz} (see also \cite{Smirnov:1994tg} and Chapter~9 
of \cite{Smirnov:2012gma}) by a simple formula with a summation over subgraphs which include all the large masses and whose connectivity components are one-particle-irreducible with respect to the lines with small masses.
Let us also mention that, for a general limit defined by treating some parameters such as kinematic invariants 
and masses as small, one can use the universal strategy of expansion by regions~\cite{Beneke:1997zp} (see also \cite{Smirnov:1999bza,Smirnov:2002pj,Smirnov:2012gma,Smirnov:2021dkb}). To do this, one can apply the public computer
code {\tt asy} \cite{Pak:2010pt,Jantzen:2012mw} (also available with the {\tt FIESTA5} distribution 
package \cite{Smirnov:2021rhf}) based on the geometry of polytopes associated with the two basic functions in the
Feynman parametric representation.  
 
The goal of this letter is to present a setup to analytically evaluate many terms within the 
large mass expansion. This setup has been developed in the framework of a 
project\footnote{M.~Bonetti and L.~Tancredi, to appear.} on the evaluation of 
two-loop form factors for the process $H \to ggg$, where the Higgs boson couples to the quarks through a pair of
massive vector bosons $V$, where $V$ is either $W^{\pm}$ or $Z$. This project has been frozen, for some reasons.
I believe that it will be sooner or later completed with the help of my setup. Moreover, I believe that the setup could be applied also in other similar calculations.
Applying the expansion in inverse powers of $m_t^2$ which is the biggest parameter in the problem 
looks natural because evaluating the corresponding Feynman integrals analytically is a 
very complicated problem.

The large mass expansion can be applied either to each of the integral involved in the calculation, or,
to the corresponding master integrals of a given family. The contribution of each of the corresponding 
subgraphs belongs to a new family of Feynman integrals, with a new set of propagators and numerators.
These integrals can be expressed, via an integration by parts (IBP) reduction~\cite{Chetyrkin:1981qh}, to the corresponding
master integrals. All the master integrals in all the contributions are considerably simpler than
the master integrals of the initial family. As it will be explained later, in the case of this project,
there are only eight ingredients appearing in the expansion.  
Six of these ingredients can be evaluated 
in terms of gamma functions at a general value of the dimensional regularization parameter, $d=4-2\ep$, 
and two of them can be evaluated in terms of multiple polylogarithms in an $\ep$ expansion up to the desired 
weight four. A technical problem at this point is to reduce integrals appearing in the expansion
to the corresponding master integrals. It turns out that, with the increase of the order of expansion,
the IBP reduction gets very complicated and time needed for such a reduction essentially increases.
It turns out, however, that this increase of complexity can be damped when applying the combination of
{\tt FIRE} and {\tt LiteRed}~\cite{Lee:2012cn,Lee:2013mka} and using the possibility to construct 
explicit analytical reduction rules with {\tt LiteRed}.

In the next Section, I present details of application of the large mass expansion to the Feynman
integrals which appear in two loops in the project on $H \to ggg$ mentioned above. I will then discuss 
some other possible improvements as well as some accompanying technical problems in Conclusion.
 
\section{Applying the large mass expansion}

There are six families of integrals appearing in two loops for the process $H \to ggg$.
They correspond to the graphs shown in Fig.~1.
\begin{figure}[ht]
\begin{center}
\includegraphics[scale=1.]{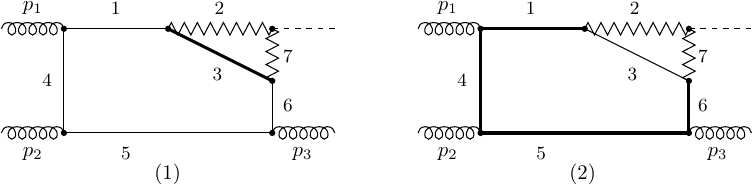}
\vspace*{2mm}
\includegraphics[scale=1.]{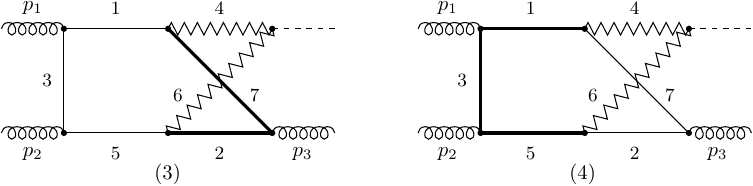}
\vspace*{2mm}
\includegraphics[scale=1.]{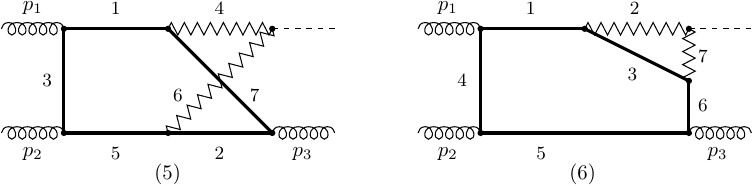}
\caption{Feynman graphs associated with the six families of integrals under consideration.  
Thick lines correspond to $m_t$ and zigzag lines correspond to $m_W$, the dashed external
line corresponds to $m_H$. All the external momenta $p_i$ are incoming.}
\label{plaBha}       
\end{center}
\end{figure} 
The general integral of each of these families takes the form
\begin{equation}
G_{a_1,a_2,\ldots,a_9} =
\int\int \frac{\dr^d k \, \dr^d l}{\prod_{i=1}^9 D_i^{a_i}}\;, 
\nn
\end{equation}
where the corresponding six sets of the propagators are 
\begin{align}
&\{k^2, l^2 - m_W^2, (k - l)^2 - m_t^2, (k - p_1)^2, (k - p_1 - p_2)^2,  (k - p_1 - p_2 - 
   p_3)^2,  \nn \\&
   (l - p_1 - p_2 - p_3)^2 -m_W^2, 
  (l - p_1)^2, (l - p_1 - p_2)^2\}\,,&
\\ 
&\{k^2 - m_t^2,  l^2 - m_W^2, (k - l)^2, (k - p_1)^2 - m_t^2, (k - p_1 - p_2)^2 - 
  m_t^2, \nn \\&(k - p_1 - p_2 - p_3)^2 - m_t^2,  (l - p_1 - p_2 - p_3)^2 - m_W^2, (l - p_1)^2, (l - p_1 - p_2)^2\}\,,
\\ 
&\{k^2, (k - l)^2 - m_t^2, (k - p_1)^2, (l + p_3)^2-m_W^2, (k - p_1 - 
     p_2)^2,   \nn \\&(l - p_1 - p_2)^2-m_W^2,(k - l - p_3)^2-m_t^2 , (l - p_1)^2 -m_W^2, (k - p_1 - p_3)^2\}\,,
\\ 
&\{k^2 - m_t^2, (k - l)^2, (k - p_1)^2 - m_t^2,  (l + p_3)^2-m_W^2, 
 (k - p_1 - p_2)^2 - m_t^2, \nn \\& (l - p_1 - p_2)^2-m_W^2,  (k - l - p_3)^2, (l - p_1)^2-m_W^2, (k - p_1 - p_3)^2\}\,,
\\ 
&\{k^2 - m_t^2, (k - l)^2 - m_t^2, (k - p_1)^2 - m_t^2, (l + p_3)^2-m_W^2 , 
(k - p_1 - p_2)^2 - m_t^2,\nn \\&  (l - p_1 - p_2)^2 -m_W^2,  (k - l - p_3)^2 - m_t^2, 
(l - p_1)^2 -m_W^2 , (k - p_1 - p_3)^2\}\,,
\\ 
& \{k^2 - m_t^2, 
 l^2 - m_W^2, (k - l)^2 - m_t^2, (k - p_1)^2 - m_t^2, (k - p_1 - p_2)^2 - m_t^2, \nn \\&
 (k - p_1 - p_2 - p_3)^2 - m_t^2,  (l - p_1 - p_2 - p_3)^2 - m_W^2, (l - p_1)^2, (l - p_1 - p_2)^2\} \,.
\end{align}
 
The last two of the nine indices can be only non-positive so that they stand for numerators. 
Let us imply the case $V=W$, for definiteness. 
Let us apply the large mass expansion in the limit $m_t \to \infty$, in spite of the fact that three end-points are on the light cone and there are massless particles. Theoretically, there is no mathematical proof of the large mass expansion in such situations but experience shows that it works.   
The mass of the top quark is bigger that the other kinematical parameters but it is not much bigger so that
we need a setup which will enable us to go to higher terms of the large mass expansion.

For Family~1, the subgraphs contributing to the expansion are
$\gamma_0=\Gamma$ (the graph itself), $\gamma_1=\{1,3,4,5,6\},\; \gamma_2=\{2,3,7\}$ and $\gamma_3=\{3\}$.
For Families~2 and~6, these are $\gamma_0=\Gamma$ and $\gamma_1=\{1,3,4,5,6\}$.
For Family~3, these are $\gamma_0=\Gamma,\; \gamma_1 = \{1, 2, 3, 5, 7\},\;
\gamma_2 = \{2, 4, 6, 7\}, \gamma_3 = \{2, 7\}$.
For Families~4 and~5, these are $\gamma_0=\Gamma$, and $\gamma_1 = \{1, 2, 3, 5, 7\}$.
 
The expansion in the large limit $m_t\to \infty$ can be described equivalently in the
language of regions which looks here preferable from the technical point of view.
For Families~1 and~3, the relevant regions corresponding to the above mentioned subgraphs are 
$\gamma_0$: $k$ and $l$ large; $\gamma_1$: $k$ large, $l$ small; $\gamma_2$: $k$ small, $l$ large; 
$\gamma_3$: $k$ and $l$ small. For the rest of the families, these are 
$\gamma_0$: $k$ and $l$ large and $\gamma_1$: $k$ large, $l$ small.
In particular, the contribution of $\gamma_0$ is given as a Taylor expansion of the integrand in
$m_W, p_1, p_2, p_3$. To write down the contribution, we make the replacements $m_W^2 \to \rho^2 m_W^2, p_1 \to \rho p_1, p_2 \to \rho p_2, p_3 \to \rho p_3$, pull out an overall power of $\rho$ and then perform a Taylor expansion in $\rho$ at $\rho=0$. Odd powers of $\rho$ should give zero results,
and this is a useful check. Finally, $\rho$ is set to 1. The other contributions are similarly constructed:
all the parameters $m_W, p_1, p_2, p_3$ as well as the small loop momenta (momentum) for a given region are multiplied 
by $\rho$.  
 
Each contribution generates a linear combination of Feynman
integrals in every order of the large mass expansion, Even if such an integral can be evaluated in terms of $\Gamma$ functions at general d, there are cumbersome numerators so that it is more effective to use immediately an IBP reduction and then it will be enough to evaluate only the corresponding master integral(s). For example, for $\gamma_0$,
the resulting Feynman integrals are vacuum integrals with one non-zero mass.
In particular, these are propagators/numerators of the four families of integrals which arise 
in the expansion in the four contributions $\gamma_0,\gamma_1,\gamma_2,\gamma_3$ for Family~1:
\begin{align}
&\{ (k - l)^2 - m_t^2, k^2, l^2, p_1\cdot k, p_1 \cdot l, p_2 \cdot k, p_2 \cdot l, p_3 \cdot k, p_3 \cdot l\}\,,&
\\ 
&\{k^2 - m_t^2, l^2 - m_W^2, (l - p_1 - p_2 - p_3)^2 - m_W^2, k \cdot  l, p_1 \cdot k, p_1\cdot  l, 
\nn \\& p_2\cdot  k, p_2 \cdot l, p_3\cdot  k  \}\,,
\\ 
&\{k^2, (k - p_1)^2, (k - p_1 - p_2)^2, (k - p_1 - p_2 - p_3)^2, l^2 - m_t^2, k \cdot l, 
p_1 \cdot l, \nn \\& p_2\cdot l, p_3\cdot l  \}\,,
\\ 
&\{ k^2, l^2 - m_W^2, (k - l)^2, (k - p_1)^2, (k - p_1 - p_2)^2, (k - p_1 - p_2 - p_3)^2, 
 \nn \\&  (l - p_1 - p_2 - p_3)^2-m_W^2 , (l - p_1)^2, (l - p_1 - p_2)^2\}\,.
\end{align}
The indices $a_i$ which can be positive are $i=1,2,3$ for the first two auxiliary subfamilies, 
$i=1,\ldots,5$ for the third subfamily and $i=1,2,4,5,6,7$ for the fourth subfamily.

For each of the contributions to the large mass expansion of integrals of all the six families, the IBP reduction is much simpler that the reduction of initial integrals. However, if we want to obtain many terms of the expansion,
the IBP reduction gets complicated because of the increase of the absolute values of the indices.
Still it is necessary to have the possibility to evaluate many terms indeed because we are oriented at the above mentioned physical problem and because the mass $m_t$ is not essentially bigger with respect to the other parameters.

It turns out that the package {\tt LiteRed} can help in this situation. The point is that, for all the subfamilies originated from the large mass expansion of integrals of the six families under consideration,
it is possible to construct explicit analytic rules for the reduction in {\em all} the corresponding sectors.
Such rules are similar in character to analytic recurrent relations obtained when solving IBP relations `by hand`
in many calculations in the period between the discovery of the IBP method~\cite{Chetyrkin:1981qh} and the appearance of
the first computer program to solve IBP relations.
These rules are constructed with the {\tt LiteRed} command {\tt SolvejSector}. It is enough to restrict the time for the command by one minute.
After running {\tt FIRE} ~\cite{Smirnov:2014hma,Smirnov:2019qkx} and taking into account these rules, the IBP reduction of the integrals
in the large mass expansion gets much faster and it becomes possible to go to higher order in $1/m_t^2$.
  
We encounter only eight different ingredients 
in all the contributions to the large mass expansion of integrals of all the six families.
Only one of them is two-loop: this is the two-loop vacuum integral with the masses $\{m_t,0,0\}$.
All the others ingredients are one-loop integrals.
Six of the ingredients are expressed in terms of gamma functions at general $d$: the one-loop vacuum integrals with the mass $m_t$ or $m_W$, the two-loop vacuum integral with the masses $\{m_t,0,0\}$, the one-loop massless propagator integrals with the external momentum squared $s,t$ or $m_H^2$. One more ingredient is the one-loop propagator integral with the two masses $m_W$, the external momentum squared $m_H^2$ and the indices $\{2,1\}$. Finally, we have the massless box integral with three end-points on the light cone and one end-point at $m_H^2$. Although it is clear that
results for them can be found somewhere in published papers, I evaluated these integrals using the method of
differential equations~\cite{Kotikov:1990kg,Gehrmann:1999as} with canonical bases~\cite{Henn:2013pwa}.
Analytic results for these `expansion master integrals' can be found in the files attached to the paper. I am also attaching
sample files with constructing terms of the large mass expansion, together with some useful {\tt Mathematica} 
auxiliary commands.

Theoretically, any order of expansion can be evaluated within this setup in terms of the basic ingredients with coefficients
which are rational in the kinematic invariants and $d$. Practically, some technical complications arise.
To illustrate them, let us consider the large mass expansion of the integral $G_{1,\ldots,1,0,0}$ of the first family.
It turns out that the contribution of the subgraph $\gamma_2$ is most complicated when going to higher orders
because the corresponding results become most cumbersome. 
At the order $1/(m_t^2)^n$, the corresponding contribution is written as a linear combination of 
$G_{1,1,1,1,1+2n,-2n,0,0,0}$ and many other integrals with less deviations from the corner point $\{1,\ldots,1,0,0\}$.
Each of these  integrals can be IBP-reduced to a linear combination of four expansion master integrals with 
rational coefficients which become cumbersome with the growth of $n$. For example, for $2n=24$, the file with results for the contribution to the expansion 
is with around 10 gb. It is not easy to handle it, i.e. to expand resulting expressions in $\ep$ and this problem with
$2(n+1)$ will only be more serious. 

One can switch to numerical evaluation at given values of kinematic invariants. For example, for this contribution of 
$\gamma_2$, turning to this mode in the IBP reduction allows to go to  $2n=30$ with results less than 1 gb.
Then it is reasonable to turn to calculations within modular arithmetic and a subsequent rational reconstruction
\cite{Peraro:2016wsq,Peraro:2019svx,Smirnov:2019qkx,Klappert:2019emp,Klappert:2020aqs,Belitsky:2023qho}.
If all the kinematic parameters are set to given values then this is a reconstruction of a rational function of $d$.
In the case of several general kinematic parameters are general one can use the latest variant \cite{Belitsky:2023qho}
of rational reconstruction based on balanced relations. Still in the project $H \to ggg$ at which this letter is oriented
the complication connected with big expressions in results arises in higher orders of the large mass
expansion so that the evaluation at fixed values can be more relevant.
On the other hand, it can happen that the order $1/(m_t^2)^{12}$ which is already accessible will be quite enough for
qualitative estimates.

\section{Conclusion}

Limits typical of Euclidean space are much simpler than limits typical of Minkowski space in various respects.
In particular, contributions of subgraphs/regions can naturally be represented as Feynman integrals also with quadratic
propagators with a standard IBP reduction.
The crucial point of the simplifications described is the possibility to construct explicit analytic rules
for an IBP reduction in all the sectors for all the subfamilies of integrals appearing in the large mass expansion.
Then the application of these rules by {\tt FIRE} and {\tt LiteRed} enables us to go to higher orders of
the expansion in $1/m_t^2$. 

In fact, one could try to derive simpler explicit reduction rules, like this was done
for massless four-loop propagators. The corresponding master integrals were evaluated in \cite{Baikov:2010hf}
and, to weight twelve, in \cite{Lee:2011jt}. It turns out that, for each of the families of these integrals, 
it is possible to construct explicit analytic rules with {\tt LiteRed}. However, the authors of~\cite{Ruijl:2017cxj}
were not satisfied by such a solution of IBP relations and produced a `hand-guided computer program'~\cite{Ruijl:2017cxj}
named {\tt Forcer} which is similar to its three-loop prototype {\tt MINCER} \cite{Gorishnii:1989gt,Larin:1991fz}.
{\tt Forcer} is certainly more powerful than the combination  {\tt FIRE}$+${\tt LiteRed}
for IBP reduction of massless four-loop propagators.
One can hope that it will be possible to construct a similar hand-guided computer program for 
integrals appearing in the large mass expansion of integrals of the six families of integrals
discussed in this paper. 
I also believe that, in the case of other two-loop four-point Feynman integrals
considered in a limit typical of Euclidean space, explicit reduction rules can exist and, alternatively,
it can be possible to constructed similar hand-guided computer programs for IBP reduction of integrals in expansion.

\vspace{0.2 cm}
{\em Acknowledgments.}
The work was supported by the Russian Science Foundation, agreement no. 21-71-30003.
I am grateful to Marco Bonetti and Ben Ruijl for discussions. 
 
\biboptions{longnamesfirst,sort&compress}
\bibliographystyle{elsarticle-num}
\bibliography{lmte}

\end{document}